\documentclass{article}

\usepackage{float,indentfirst}
\usepackage{amssymb,amsfonts,amsmath}
\usepackage{color}

\def\be{\begin{eqnarray}}
\def\ee{\end{eqnarray}}
\def\nn{\nonumber}

\def\p{\partial}
\def\tr{{\rm tr}\,}
\def\Tr{{\rm Tr}\,}

\makeatletter
\renewcommand*{\@cite@ofmt}{\bfseries\hbox}
\makeatother

\definecolor{red}{rgb}{1,0,0}
\definecolor{orange}{rgb}{1,0.5,0}
\definecolor{violet}{rgb}{0.7,0,1}

\hoffset= - 1.0in         

\begin{document}

\title{\vspace{-.5cm}{\Large {\bf  Sum rules for characters
from character-preservation property of matrix models}\vspace{.2cm}}
\author{
{\bf A.Mironov$^{a,b,c}$}\footnote{mironov@lpi.ru; mironov@itep.ru}\ \ and
\ {\bf A.Morozov$^{b,c}$}\thanks{morozov@itep.ru}}
\date{ }
}

\maketitle

\vspace{-5.5cm}

\begin{center}
\hfill FIAN/TD-12/18\\
\hfill IITP/TH-12/18\\
\hfill ITEP/TH-18/18
\end{center}

\vspace{3.3cm}

\begin{center}
$^a$ {\small {\it Lebedev Physics Institute, Moscow 119991, Russia}}\\
$^b$ {\small {\it ITEP, Moscow 117218, Russia}}\\
$^c$ {\small {\it Institute for Information Transmission Problems, Moscow 127994, Russia}}

\end{center}

\vspace{.5cm}

\begin{abstract}
One of the main features of eigenvalue matrix models
is that the averages of characters are again characters,
what can be considered as a far-going generalization of
the Fourier transform property of Gaussian exponential.
This is true for the standard Hermitian and unitary (trigonometric) matrix models
and for their various deformations, classical and quantum ones.
Arising explicit formulas for the partition functions are very efficient
for practical computer calculations.
However, to handle them theoretically,
one needs to tame remaining finite sums over representations of a given size,
which turns into an interesting conceptual problem.
Already the semicircle distribution in the large-$N$ limit
implies interesting combinatorial sum rules for characters.
We describe also implications to $W$-representations,
including a character decomposition of cut-and-join operators,
which unexpectedly involves only single-hook diagrams
and also requires non-trivial summation identities.
\end{abstract}

\bigskip

\bigskip

\section{Introduction}

As emphasized quite recently in \cite{HMAMOsol}
following the consideration in \cite{IMMrainbow},
a key feature of the Gaussian measures is that really nice
are the averages of characters. In particular, in the Hermitian matrix model \cite{HerMAMO, UFN3},
the average of a character (Schur function) $\chi_R[M]$, which is
a function of eigenvalues of the matrix variable $M$,
\be
\boxed{
\Big<\chi_R[M]\Big> =
\chi_R\{N\} \cdot \frac{\chi_R\{\delta_{k,2}\}}{\chi_R\{\delta_{k,1}\}}
}
\label{Hermave}
\ee
is again a character, actually, a dimensions $D_R(N)= \chi_R[I_N]=\chi_R\{N\}$ of the representation $R$ of $gl_N$. 
For monomial non-Gaussian measures like $e^{\tr M^s}dM$
and appropriate choice of integration contour, the coefficient
contains $\chi_R\{\delta_{k,s}\}$ \cite{PSh}.
In the present paper, we concentrate on the case of
Gaussian measures with $s=2$.
We use the square and curled brackets in order to denote the character as a (symmetric) function of matrix eigenvalues (the first Weyl formula) and as a function of time variables $\tr M^k$ (the second Weyl formula) accordingly. The matrix eigenvalues are often called Miwa variables within this context.

As an example of (\ref{Hermave}), $\Big<\Tr M^2\Big>=N^2$ and $\Big<(\Tr M)^2\Big> = N$ imply that
$\left<\frac{\Tr M^2 \pm (\Tr M)^2}{2}\right> = \frac{N(N\pm 1)}{2}$,
which are dimensions of the symmetric and antisymmetric representations [2] and [1,1].
The only non-trivial ingredient of the theory is the coefficient,
which is actually a ratio of characters at two peculiar points in the
space of time-variables, $p_k = \delta_{k,2}$ and $p_k=\delta_{k,1}$.
It is this $R$-dependent coefficient, which makes the matrix model somewhat non-trivial.
At the same time, the character-preserving property can be considered
as a defining feature of the Gaussian measures, and can serve as a key
for the definition of various deformations of the Hermitian model
defined by change of the Schur functions to other systems of orthogonal symmetric functions
\cite{Mac,MPS}.

In more detail, the partition function of the Hermitian matrix model \cite{HerMAMO,UFN3},
i.e. the Gaussian average over $N\times N$ Hermitian matrices $M$, can be decomposed into
a sum over all Young diagrams $R$:
\be
Z_N\{p\}
= \frac{\mu^{N^2/2}}{ {\rm Vol}_{U_N}}\int dM \exp\left(-\frac{\mu}{2}\Tr M^2 +
\sum_k \frac{p_k}{k}\,\Tr M^k\right)
=\left<\exp \left(\sum_k \frac{p_k}{k}\,\Tr M^k\right)\right>=
\nn \\
\boxed{
= \sum_\Delta Z_\Delta(N) \cdot p_\Delta
= \sum_R \chi_R\{p\}\cdot\Big<\chi_R\{\Tr M^k\}\Big>
= \sum_R \frac{ \chi_R\{\delta_{k,2}\}\cdot\chi_R\{N\}\cdot\chi_R\{p\} }
{\mu^{|R|/2}\chi_R\{\delta_{k,1}\}}
}
\label{chardeco}
\ee
From now on, for the sake of simplicity, we put $\mu=1$, it can be easily restored by dimensional analysis. In (\ref{chardeco}), the character
$\chi_R\{p\}$ is a polynomial of the time variables $p_k$, labeled by the Young diagram $R$ :
\be
\chi_R\{p\} = \sum_{\Delta\vdash |R|} \frac{\psi_{_{R,\Delta}}}{z_{_\Delta}}\cdot p_\Delta
= \sum_{\Delta\vdash |R|} d_R \cdot\varphi_{R,\Delta} \cdot p_\Delta
\label{chapdeco}
\ee
where, for the Young diagram $\Delta$ parameterized in one of the two ways
$\Delta=[\delta_1,\delta_2,\ldots]=[1^{m_1},2^{m_2},\ldots]$,
\be
p_{_\Delta} = \prod_i p_{\delta_i}, \ \ \ \ \ \ \
z_{_\Delta} = \prod_k(k!)^{m_k}m_k!
\ee
and $\psi_{_{R,\Delta}}$ is the symmetric group character, and $\varphi_{R,\Delta}$, its sometimes more convenient rescaled.
The symbol $\vdash$ is used in Hurwitz theory to denote restricting to diagrams
of a given size, and $|R|$ is the number of boxes in the Young diagram $R$.
The $N$-dependence at the r.h.s. of (\ref{chardeco}) comes entirely from the dimensions
$\chi_R\{N\}=D_R(N)$, which are the values of characters at the locus where all $p_k=N$.
Standing in the denominator of (\ref{chardeco}) is an important
representation theory quantity often denoted by $d_R$:
\be
d_R = \chi_R\{\delta_{k,1}\} = {1\over |R|!}{\prod_{i<j}(r_i-i-r_j+j)\over \prod (l_{_R}+r_i-i)!}=\prod_{\hbox{all boxes of }R}{1\over\hbox{hook length}}
\ee
where $l_{_R}$ is the number of lines in the Young diagram $R$.
The ``classical'' tool to deal with the characters is the Cauchy formula
\be
\sum_R  \chi_R\{p\}\chi_R\{\bar p\} = \exp\left(\sum_k \frac{p_k\bar p_k}{k}\right)
\label{Cauchy}
\ee
and it turns sufficient to solve surprisingly many problems, involving characters.
But not all, as we immediately see in what follows, and as is already clear from
(\ref{chardeco}), where the summand is still of the second-order in characters, but
not just bilinear: it contains one extra character in the numerator
and another one, $d_R$, in the denominator.

In the particular case of $N=2$, the only relevant are characters of the symmetric
representations $R=[r]$, which, when expressed through the two eigenvalues of $M$,
are just $\chi_{[r]}[M]=\sum_{i=0}^r m_1^im_2^{r-i}$.
Then the l.h.s. of (\ref{Hermave}) is
\be
\Big<\chi_{[r]}[M]\Big>^{N=2} = \frac{\int\int \left(\sum_{i=0}^r m_1^im_2^{r-i}\right)
(m_1-m_2)^2 e^{-m_1^2/2-m_2^2/2}dm_1dm_2}
{\int\int (m_1-m_2)^2 e^{-m_1^2/2-m_2^2/2}dm_1dm_2} =  \left\{\begin{array}{cl}
(r+1)!!& \hbox{for }r\hbox{ even}\\ &\\
0&\hbox{otherwise}
\end{array}
\right.
\label{lhsHermave}
\ee
For symmetric representations, the special characters can be easily obtained
from the Cauchy formula at  $\bar p_k=x^k$:
\be
\sum_r x^r\chi_{[r]}\{p_k\} = e^{\sum_k \frac{p_kx^k}{k}}
\ \ \Longrightarrow \ \ \sum_r x^r\chi_{[r]}\{\delta_{k,m}\} = \sum_j \frac{x^{mj}}{j! \cdot m^j}
\ee
i.e. $\chi_{[r]}\{\delta_{k,1}\} = \frac{1}{r!}$, and
$\chi_{[r]}\{\delta_{k,2}\} = \frac{\delta_{r,{\rm even}}}{2^{r/2}(r/2)!}
= \frac{\delta_{r,{\rm even}}}{r!!}$.
Thus, for  the r.h.s.  of (\ref{Hermave}), we get
\be
\chi_{[r]}(2)\cdot\frac{\chi_{[r]}\{\delta_{k,2}\}}{\chi_{[r]}\{\delta_{k,1}\}}
= (r+1)\cdot\frac{r!}{r!!} = (r+1)!!
\ee
which coincides with (\ref{lhsHermave}).

The need for a character in the denominator of (\ref{chardeco}) becomes nearly obvious,
if we extend the Hermitian model to the rectangular complex one \cite{MMMM},
where the variable $M$ is $N_1\times N_2$ rectangular matrix
not obligatory square.
Then the average should depend on two parameters $N_1$ and $N_2$ in symmetric way.
Thus one expects, and gets two characters $\chi_R\{N_1\}\chi_R\{N_2\}$
in the numerator \cite{HMAMOsol}:
\be
\boxed{
\frac{1}{ {\rm Vol}_{U_N}} \int dM e^{-\Tr MM^\dagger} \chi_R[MM^\dagger]
= \frac{\chi_R\{N_1\}\chi_R\{N_2\}}{\chi_R\{\delta_{k,1}\}}
= \frac{D_R(N_1)D_R(N_2)}{\chi_R\{\delta_{k,1}\}}
}
\label{chardecomplex}
\ee
and therefore there should be one in the denominator as well in order to balance
the number of characters at the l.h.s. and the r.h.s.

\bigskip

In this paper, we review character sum rules arising from the large-$N$ expansion,
and describe a more general approach based on use of cut-and-join operators
introduced in \cite{MMN} and playing a big role in the theory of Hurwitz tau-functions.
We also comment on extension to various models from the unitary (trigonometric, Chern-Simons, ... family,
including torus knot and MacMahon models.

\section{Sum rules from comparison to Harer-Zagier formula}

Besides (\ref{chardeco}),
there are  other explicit generating functions
like the Harer-Zagier formula \cite{HZ}
\be
\sum_n \frac{z^{2m}}{(2m-1)!!}\cdot\Big<\Tr M^{2m}\Big>
= \frac{1}{2z^2}\left(\left(\frac{1+z^2}{1-z^2}\right)^N-1\right)
\label{HZ1}
\ee
(for increasingly sophisticated multi-trace generalizations see \cite{MShonHZ}).

Note that the same coefficient $Z_{[2m]}$ in front of $p_{[2m]}=p_{2m}$ can be read off from (\ref{chardeco}), and
is provided by the sum
\be
Z_{[2m]} = \frac{1}{2m}\Big<\Tr M^{2m}\Big> =
\sum_{R\vdash 2m} \frac{\chi_R\{N\}\chi_R\{\delta_{k,2m}\}\chi_R\{\delta_{k,2}\}}
{\chi_R\{\delta_{k,1}\}}
\label{Z2m}
\ee
Now, substituting (\ref{Z2m}) into (\ref{HZ1}), we obtain a non-trivial
sum rule for characters:
\be\label{1}
\boxed{\boxed{
\sum_{R }\frac{2|R|z^{|R|+2}}{(|R|-1)!!}\cdot
\frac{ \chi_R\{\delta_{k,|R|}\}\chi_R\{\delta_{k,2}\}}
{\chi_R\{\delta_{k,1}\}}\cdot D_R(N)
=  \left(\frac{1+z^2}{1-z^2}\right)^N-1-2Nz^2
}}
\ee
This formula can be definitely also derived from combinatorics, using that
\be
\chi_R\{\delta_{k,|R|}\}=\left\{\begin{array}{cl}{(-1)^d\over r}&\hbox{if }R=[r-d,1^d]\\ &\\
0&\hbox{otherwise}\end{array}\right.
\ee
i.e. only the hook Young diagrams $R$ contribute, moreover, since $\chi_R\{\delta_{k,2}\}$ is non-zero only at even $|R|$, we can parameterize the hook diagrams as $R=[r-d,1^d]$ at even $r$. For these Young diagrams,
\be
\chi_R\{\delta_{k,2}\}={\psi_{[r-d,1^d],[2^{r/2}]}\over z_{[2^{r/2}]}}={(-1)^{d+\mathfrak{i}_d}C_{r/2-1}^{\mathfrak{i}_d}\over r!!}
\\
{D_R(N)\over\chi_R\{\delta_{k,1}\}}=\prod_{i,j\in R}(N+j-i)={(N+r-d-1)!\over (N-d-1)!}
\ee
where $C_n^m$ are the binomial coefficients, and $\mathfrak{i}_d$ denotes the integer part of $d/2$.
However, calculating the l.h.s. of (\ref{1}) does require additional summations over $r$ and $d$, and the derivation of (\ref{1}) becomes not that immediate.

Actually, it is easy to check that
\be
Z_N\{p\} = \sum_{m=2}^\infty\left\{
\frac{p_2^m}{2^m m!}\cdot\prod_{i=1}^m(N^2+2i-2)
+  \frac{p_1^m}{2^mm!}\cdot N^m
+ \ldots \right\}
\nn \\
+ \sum_{m=1}^\infty\frac{2^{2m-1}p_{2m}}{m} \cdot
\frac{ \Gamma(m+1/2)}{ \Gamma(1/2)}\cdot
\sum_{k=0}^{m/2} \frac{N^{m+1-2k}}{(2k+1)!(m-2k)!}\cdot\frac{\xi_k^{(1)}}{m+1}
+ \nn \\
+\sum_{m=2}^\infty \sum_{k=0}^{m-1}
\frac{2^{2(m-1)}p_{2k+1}p_{2m-2k-1}}{m\cdot k!(m-k-1)!}
\frac{\Gamma(k+1/2)\Gamma(m-k-1/2)}{\Gamma(1/2)^2}\Big(N^m + \ldots\Big)+\ldots
\ee
where we explicitly write down a few typical terms in the expansion at the r.h.s.
Expansion coefficients here are:
{\footnotesize
\be
\xi^{(1)}_0=1, \ \ \ \
\xi^{(1)}_1=\frac{m+1}{2}, \ \ \ \
\xi^{(1)}_2=\frac{(m+1)(5m-2)}{12}, \ \ \ \
\xi^{(1)}_3=\frac{(m+1)(35m^2-77m+12)}{72}, \ \ \ \
\xi^{(1)}_4=\frac{(m+1)(175m^3-945m^2+1094m-72)}{240}, \ \ \ \
\ldots
\nn
\ee
}
The simplest here is the $p_1^{2m}$ term: it comes from the contribution
$\chi_R\{\delta_{k,1}\}p_1^{|R|}$ to $\chi_R\{p\}$,
and the coefficient cancels the denominator (\ref{chardeco})
so that the remaining sum is calculated with the help of the Cauchy formula,
\be
\sum_R\chi_R(N)\chi_R\{\delta_{k,2}\}p_1^{|R|} = e^{Np_1^2/2}=
\sum_m \frac{p_1^{2m}}{2^mm!}\cdot N^m
\ee

One can say that the $p_2^m$ term is also easy,
since it can be obtained by differentiating the Gaussian integral:
$(-2\p_\mu)^m \mu^{-N^2/2} = \frac{N^2(N^2+2)\ldots(N^2+2m-2)}{\mu^m} \mu^{-N^2/2}$.
However, from the point of view of the sum (\ref{chardeco}),
this is already a non-trivial sum rule
for the dimensions $D_N=\chi_R(N)$:
\be
\boxed{\boxed{
\sum_{R\vdash m}\frac{\chi_R\{\delta_{k,2}\}^2}{\chi_R\{\delta_{k,1}\}}\cdot
D_R(N)
=\frac{1}{2^m m!}\cdot\prod_{i=1}^m(N^2+2i-2)
}}\label{n22}
\ee
For $m=1$, it is still trivial:
$\frac{(1/2)^2}{1/2}\cdot\frac{N(N+1)}{2} +\frac{(1/2)^2}{1/2}\cdot\frac{N(N-1)}{2}=\frac{N^2}{2}$,
but already for $m=2$ it is not:
$\frac{(1/8)^2}{1/24}\cdot\frac{N(N+1)(N+2)(N+3)}{24}
+ \frac{(-1/8)^2}{1/8}\cdot \frac{(N-1)N(N+1)(N+2)}{8}
+\frac{(1/4)^2}{1/12}\cdot \frac{(N-1)N^2(N+1)}{12}
+\frac{(-1/8)^2}{1/8}\cdot \frac{(N+1)N(N-1)(N-2)}{8}
+\frac{(1/8)^2}{1/24}\cdot \frac{N(N-1)(N-2)(N-3)}{24}
= \frac{2}{64}(N^4+11N^2) + \frac{2}{64}(N^4-N^2)+ \frac{1}{16}(N^4-N^2)
= \frac{N^2(N^2+2)}{2^2\cdot 2!}$.
The natural question is how one can handle all the variety of the sum rules for characters,
which arise in this way.

\bigskip

The question becomes even more interesting, because the result (\ref{chardeco})
of \cite{HMAMOsol}
possesses wide generalizations: to various deformations ($q-,t-$ of \cite{5dMAMO,DIM}
and many other)
of matrix models \cite{MPS} and, in another direction,
to Aristotelian and other tensor models \cite{MKR1,tenmod,Rang}.
At the same time, already at the Hermitian matrix model level, it has important applications,
the currently fashionable ones being related to localization formulas \cite{loca,LMNS,Pest,Zabz}
in conformally invariant supersymmetric field theories, which reduce
perturbative contributions to certain correlators to  those in the Gaussian matrix model averages
\cite{N2mamo}.

\section{Harer-Zagier formula and planar limit}

Let us restore the $\mu$-dependence in (\ref{chardeco}), and consider the planar large-$N$ ('t Hooft) limit $N\to \infty$, $\nu:=N/\mu=$fixed. Then, the model is described by the semicircle distribution of eigenvalues
\be\label{sc}
\rho(z) := \Big< \Tr \delta(M-z\cdot I)\,dz \Big> = \sqrt{4\nu-z^2}\,dz +O(N^{-2})
\ee
This means, in particular, that, in the large-$N$ approximation  (for the sake of simplicity, we put $\nu=1/4$),
\be
\!\!\!\!\!\!\!\Big<\Tr M^{2m}\Big> \longrightarrow  {2\over \pi} \int_{-1}^1 z^{2m}\sqrt{1-z^2}dz =
\int_0^\pi \sin^{2m}\!t\, \cos^2t \,dt = \frac{\pi}{2^{2m-1}}\left( \frac{(2m)!}{(m!)^2} -
\frac{(2m+2)!}{4\left((m+1)!\right)^2}\right) =
\frac{(2m-1)!!}{2^{m}(m+1)!}
\label{largeNas}
\ee
Consistency with (\ref{HZ1}) is straightforward: its r.h.s.
in 't Hooft limit is equal to
\be
\lim_{N\rightarrow\infty}
\frac{1}{ z^2}\left(\left(\frac{1+\frac{z^2}{2N}}{1-\frac{z^2}{2N}}\right)^N-1\right)
= \frac{e^{z^2}-1}{2z^2} = \sum_n \frac{z^{2n}}{(n+1)!}
\label{HZlargeNas}
\ee
and this is exactly what one gets by substituting (\ref{largeNas})
into the l.h.s. of (\ref{HZ1}).

More interesting is consistency with (\ref{chardeco}).

Of course, it is  straightforward to get the leading contribution to (\ref{chardeco}),
because the leading
large-$N$ asymptotics of $\chi_R\{N\} = d_RN^{|R|} + \ldots$ comes from
$p_1^{|R|}$ and thus is proportional to $d_R = \chi\{\delta_{k,1}\}$, which
stands in the denominator.
Thus the main large-$N$ asymptotics of (\ref{chardeco}) is  controlled by the
Cauchy formula:
\be
Z_N\{p\} = \sum_R N^{|R|}\chi_R\{p\}\chi_R\{\delta_{k,2}\} + \ldots
= \exp\left(\sum_k \frac{N^kp_2^k\delta_{k,2}}{k}\right) =
\exp \left(\frac{N^2p_2}{2}\right) = \sum_m \frac{N^{2m}}{2^mm!}\cdot p_2^m
\ee
However, this is not what we need for comparison with (\ref{HZ1}).
Indeed the relevant coefficient $Z_{[2m]}$ in front of $p_{[2m]}=p_{2m}$
is provided by the sum (\ref{Z2m}),
and if we substitute $d_RN^{|R|}$ instead of $\chi_R\{N\}$ and use the Cauchy formula,
we get just nothing, unless $m=1$.
In fact, the leading asymptotics of  $Z_{[2m]}$ is
defined by sub-leading $O(N^{|R|+1-m})$ terms in $\chi_R\{N\}$,
and, hence, the sum is not reduced to the Cauchy formula.

\section{Sum rules from genus expansion}

The simple calculation in eqs.(\ref{HZ1})-(\ref{n22}) has a lot of
generalizations, to arbitrary coefficients $Z_\Delta$.
The leading asymptotics is prescribed by the semicircle distribution and is factorized
into contributions of the symmetric (single-line) diagrams $\Delta$.
In general, one can use
the well-studied genus expansion of the Hermitian model partition function \cite{AMM1,AMM},
build from the semicircle distribution (\ref{sc})
by solving the Virasoro constraints \cite{vircor,UFN3}
(this is also known as the AMM/EO topological recursion \cite{AMM1,AMM,AMM/EO,AMMU}).
The first sum rules implied by the known resolvents from \cite{AMM1} are (here $y(z) = \sqrt{z^2-4N}$)
\be
\left<\Tr M^{2m}\right> =
2m \cdot\sum_{R\vdash 2m}
\frac{\chi_R\{N\}\chi_R\{\delta_{k,2}\}}
{\chi_R\{\delta_{k,1}\}} \cdot {\rm coeff}_{p_{2m}}\chi_R\{p\}
= 2m\cdot \sum_{R\vdash 2m} \varphi_{R,[2m]} D_R\{N\}\chi_R\{\delta_{k,2}\}
= \nn \\
= {\rm coeff}_{z^{-2m-1}}\left(\frac{z-y(z)}{2} + \frac{N}{y^5(z)} +\frac{21N(z^2+ N)}{y^{11}(z)}
+ \frac{11N(135z^4+558Nz^2+158N^2)}{y^{17}(z)} + \ldots \right)
\ee

\bigskip

\be
\left<\Tr M^{k}\,\Tr M^{2m-k}\right>
= k(2m-k)\!\! \sum_{R\vdash 2m} \!\!
\frac{\chi_R\{N\}\chi_R\{\delta_{k,2}\}}{\chi_R\{\delta_{k,1}\}}
\,{\rm coeff}_{p_kp_{2m-k}}\chi_R\{p\}
= k(2m-k)\!\! \sum_{R\vdash 2m} \!\! \varphi_{R,[2m-k,k]} D_R\{N\}\chi_R\{\delta_{k,2}\}
= \nn
\ee
\vspace{-0.3cm}
\be
= \left<\Tr M^{k}\right>\left<\Tr M^{2m-k}\right>
+ {\rm coeff}_{z_1^{-k-1}z_2^{-(2m-k)-1}}\left\{
\frac{1}{2(z_1-z_2)^2}\left(\frac{z_1z_2-4N}{y(z_1)y(z_2)}-1\right)
+ \right.
\ee
\vspace{-0.3cm}
\be
\!\!\!\!\!\!\!\!\!\!\!\!\!\!\!\!\!\!\!\!\!\!\!\!\!\!
\left.
+ {\footnotesize
 \frac{N\Big( z_1z_2(5z_1^4+4z_1^3z_2+3z_1^2z_2^2+4z_1z_2^3+5z_2^4)
+4N\Big\{z_1^4-13z_1^2z_2^2(z_1^2 +z_1z_2+z_2^2)+z_2^4\Big\}
+16N^2(-z_1^2+13z_1z_2-z_2^2)+320N^3\Big) }{y(z_1)^7y(z_2)^7}
}
+\ldots\right\}
\nn
\ee
and so on.
Of course, one can convert this into generating functions, which produces the sum rules involving the whole resolvent $\mathfrak{O}_1(z):=\left<\Tr \frac{1}{z-M}\right>$:
\be
\frac{N}{z}+
\sum_R \frac{\varphi_{R,[|R|]}|R|D_R(N) \chi_R\{\delta_{k,2}\}}{z^{|R|+1}}
= \left<\Tr \frac{1}{z-M}\right>
= \frac{z-y(z)}{2} + \frac{N}{y^5(z)} +\frac{21N(z^2+ N)}{y^{11}(z)} + \ldots
\label{1ptsumrule}
\ee
As we explained earlier, contributing to the l.h.s. are actually only the 1-hook diagrams $R$:
only they have non-vanishing $\varphi_{R,[|R|]}$.

Thus, knowledge of the resolvents immediately allows one to generate sum rules, or, to put it differently, one can express resolvents as character sums this way.

\section{Gaussian averages of exponentials (Wilson loops)}

Formulas like (\ref{1ptsumrule}) can become a little less mysterious,
if the r.h.s. is written in a somewhat different way.
In fact, such a possibility is provided by the theory of exponential correlators.

According to \cite{BreHi}, the generating function of exponentials in the Hermitian matrix model is
given by the simple integral
\be
E(s_1,\ldots,s_n) =
\left< \prod_{\alpha=1}^n \Tr e^{s_\alpha M} \right>
= \prod_{\alpha=1}^n \frac{e^{s_\alpha^2/2}}{s_\alpha}
\oint du_\alpha e^{u_\alpha s_\alpha}\left(1+\frac{s_\alpha}{u_\alpha}\right)^N
\prod_{\alpha<\beta}
\frac{(u_\alpha-u_\beta)(u_\alpha-u_\beta+s_\alpha-s_\beta)}
{(u_\alpha-u_\beta-s_\beta)(u_\alpha-u_\beta+s_\alpha)}
\ee
and it provides a short way \cite{MShonHZ} to generalization of the Harer-Zagier
formulas like (\ref{HZ1}).
It is related to resolvents by the Laplace transform,
\be
\mathfrak{O}_n(z_1,\ldots,z_n) = \left<\prod_{\alpha=1}^n\Tr \frac{1}{z_\alpha-M}\right>
= \prod_{\alpha=1}^n \int_0^\infty e^{-s_\alpha z_\alpha} E(s_1,\ldots,s_n)
\ee
For $n=1$, this gives (see also \cite[eq.(IV.1.11)]{AMM1})
\be
\!
\mathfrak{O}_1(z)=\int_0^\infty \!\!\!\!
\underbrace{e^{s^2/2}}_{\sum_k \frac{s^{2k}}{2^kk!}}\!\!\!\cdot\, e^{-sz}\ ds   \cdot
\underbrace{{\rm res}_{u=0} \left\{\frac{e^{su}}{s}\left(1+\frac{s }{u }\right)^N\right\}}_
{\sum_{i=1}^N \frac{N!}{i!(N-i)!(i-1)!}  s^{2i-2} }
= \sum_{i,k }^\infty \frac{z^{1-2k-2i}}{2^kk!}\frac{(2k+2i-2)!\,N!}{(N-i)!i!(i-1)! }
= \frac{z-y(z)}{2} + \frac{N}{y^5(z)} + \ldots
\ee
Note that expanded is the quadratic exponential, not the linear one, because $E(s)$ is treated
as a series in $s$.

Using this formula, we can rewrite the sum rule (\ref{1ptsumrule}) without $y(z)$ as
\be
\boxed{\boxed{
 \sum_R \frac{\varphi_{R,[|R|]}|R|D_R(N) \chi_R\{\delta_{k,2}\}}{z^{|R|+1}}
=- \frac{N}{z}
+ \sum_{k=0}^\infty \sum_{i=0}^{N-1} \
\frac{z^{-1-2k-2i}}{2^kk!}\frac{(2k+2i)!\,N!}{(N-i-1)!\,i!\,(i+1)! }
}}
\ee
Contributing at the both sides are only odd negative powers of $z$ beginning from $z^{-3}$.

\section{Cut-and-join operator}

Using the second relation in (\ref{chapdeco}),
\be
\chi_R\{p\}
= \sum_{\Delta\vdash |R|} d_R \cdot\varphi_{R,\Delta} \cdot p_\Delta
\label{chapdeco1}
\ee
we can trade the denominator in (\ref{chardeco}) for the character $\varphi$:
\be
Z_N\{p\} =
\sum_R \sum_{\Delta\vdash |R|} \varphi_{R,\Delta} \chi_R\{N\}\chi_R\{\delta_{k,2}\} \cdot p_\Delta
= \sum_\Delta Z_\Delta\cdot p_\Delta
\ee
and then use the fact that $\varphi$ is the eigenvalue of the
generalized cut-and-join operator \cite{MMN}
\be
\hat W_{\!_\Delta} \chi_R\{\bar p\} = \varphi_{R,\Delta}\cdot \chi_R\{\bar p\}
\label{evW}
\ee
where
\be
\hat W_{\!_\Delta} = \frac{1}{z_{_\Delta}}:\prod_i \hat D_{\delta_i}:
\label{Wops}
\ee
and
\be
\hat D_k = \Tr (M \p_{M})^k
\ee
acts on the time-variables $\bar p_k = \Tr \bar M^k$.
The normal ordering in (\ref{Wops}) implies that all the derivatives $\p_M$
stand to the right of all $M$.
Since $W_\Delta$ are ``gauge"-invariant matrix operators, and we apply them only to
gauge invariants, they can be realized as differential operators in $\bar p_k$ \cite{MMN}.

Then the coefficient in front of $p_\Delta$ in $Z_N\{p\}$ can be represented as
\be
Z_\Delta =  \sum_{R\vdash|\Delta| }  \varphi_{R,\Delta} \chi_R\{N\}\chi_R\{\delta_{k,2}\}
\ \stackrel{(\ref{evW})}{=}\ \left.\hat W_{\!_\Delta} \sum_{R\vdash|\Delta|}
\chi_R\{\bar p\}\chi_R\{\delta_{k,2}\}\right|_{\bar p_k=N}
\!\!\!\!\stackrel{(\ref{Cauchy})}{=}\
\left.\hat W_{\!_\Delta} e_{|\Delta|}^{\bar p_2/2}\right|_{\bar p_k=N}
\label{ZthrW}
\ee
i.e.
\be
\boxed{
Z_\Delta = \frac{1}{2^{|\Delta|/2}(|\Delta|/2|)!}
\left.\hat W_{\!_\Delta} \, \bar p_2^{|\Delta|/2}\right|_{\bar p_k=N}
}
\label{ZthroughW}
\ee
where $e^x_n$ denotes projection to grading $n$,
which is needed in (\ref{ZthroughW})
because the sum over Young diagrams is restricted to a given size
(note that $p_2$ has the grading degree $2$).
For example,
$\hat W_{[1^{2m}]} = \frac{1}{ (2m)!}:\hat W_{[1]}^{2m}:$
where $\hat W_{[1]} = \hat D_1= \sum_{k} k\bar p_k\p_{\bar p_k} $,
i.e.multiplies $p_2^{m}$ by $2m$.
The normal ordering means that $:\hat W_{[1]}^{2m}:$ multiplies it by $(2m)!$,
and therefore the coefficient in front of $p_{[1^{2m}]}=p_1^{2m}$ in $Z_N\{p\}$ is
\be
Z_{[1^{2m}]} = \frac{1}{2^mm!} \cdot \frac{1}{(2m)!}\cdot(2m)!\cdot N^m = \frac{N^m}{2^mm!}
\ee

Alternatively one can rewrite (\ref{ZthrW}) as
\be
Z_\Delta =  \sum_{R\vdash|\Delta| }  \varphi_{R,\Delta} \chi_R\{N\}\chi_R\{\delta_{k,2}\}
\ \stackrel{(\ref{evW})}{=}\  \left.\hat W_{\!_\Delta} \sum_{R\vdash|\Delta|}
\chi_R\{N\}\chi_R\{\bar p\} \right|_{\bar p_k=\delta_{k,2}}
\!\!\!\stackrel{(\ref{Cauchy})}{=}\  \left.\hat W_{\!_\Delta}
e_{|\Delta|}^{N\sum_{k}\frac{\bar p_k}{k}}\right|_{\bar p_k=\delta_{k,2}}
\ee
where again the index $|\Delta|$ means that one should pick up a contribution of
particular grading degree.

\section{$W$-representations in terms of characters}

Eq.(\ref{ZthroughW}) is a kind of a dual to the $W$-representation \cite{MShWrep}
of Hermitian partition function
\be
Z_N\{p\} = e^{\frac{1}{2}\hat {\cal W}_2} \cdot 1
\label{HermWrep}
\ee
which involves a close relative
\be
\hat{\cal W}_2 =
N^2p_2+Np_1^2+2N\sum_{a=1}^\infty ap_{a+2}\p_a
+\sum_{a,b=1}^\infty \Big((a+b-2)p_ap_b\p_{a+b-2} + ab\,p_{a+b+2}\p_a\p_b\Big)
\label{calW2}
\ee
of the simplest cut-and-join operator $\hat{W}_{[2]}$, \cite{GD},
\be\label{W2}
\hat{ W}_{[2]} ={1\over 2} \sum_{a,b} \Big((a+b)p_ap_b\p_{a+b} + abp_{a+b}\p_a\p_b\Big)
\ee
but in (\ref{HermWrep}) this operator  is exponentiated in contrast with (\ref{ZthroughW}).

The average of character in the $W$-representation is equivalent to action
of the differential operator $\chi_R\{\p/\p t_k\}$.
Moreover, since ${\cal W}_2$ in (\ref{HermWrep}) has non-trivial grading (+2),
only the single term
of the exponential expansion contributes to the average:
\be
\Big<\chi_R\{P_k=\Tr M^k\}\Big> \ = \ \left.
\frac{1}{2^{|R|/2}(|R|/2)!}\ \chi_R\left\{k\frac{\p}{\p p_k}\right\}\
\hat{\cal W}_2^{|R|/2} \cdot 1 \ \right|_{{\rm all}\ p_k=0}
\ee
Contributions at odd levels $|R|$ are vanishing.
At the level $|R|=2$ this gives for $R=[2]$ and $[1,1]$:
\be
\left.\frac{1}{2}\chi_R\left\{\frac{1}{k}\frac{\p}{\p p_k}\right\}
\ \hat{\cal W}_2\cdot 1\ \right|_{p=0}
= \left.\frac{1}{4}\,(\pm{2}\p_2+\p_1^2)\,(N^2p_2+Np_1^2)\ \right|_{p=0} = \frac{N(N\pm 1)}{2}
\ee
For arbitrary even $|R|$ the highest power of $N$ comes from the term
$N^{|R|}p_2^{|R|/2}$ in $\hat{\cal W}_2^{|R|/2}$, and is equal to
\be
\frac{1}{2^{|R|/2}(|R|/2)!}\ \chi_R\{\delta_{k,2}\} \left(2\frac{\p}{\p{p_2}}\right)^{|R|/2}
\!\!N^{|R|}p_2^{|R|/2}
= \chi_R\{\delta_{k,2}\}\cdot N^{|R|}
\ee
which differs by a factor $d_R=\chi_R\{\delta_{k,1}\}$ from the item
$d_R N^{|R|}$ in $D_R(N)$.

In general,
(\ref{Hermave}) implies that
\be
e^{\frac{1}{2}\hat {\cal W}_2}\cdot 1 =
\sum_R \chi_R\{p\} \cdot \chi_R\{\delta_{k,2}\}\cdot \frac{D_R(N)}{\chi_R\{\delta_{k,1}\}}
\label{Zhermans}
\ee
Similarly, for the rectangular complex matrix model
\be
e^{\hat{\cal W}_1} \cdot 1 = \sum_R \chi_R\{p\} \cdot \frac{D_R(N_1)D_R(N_2)}{\chi_R\{\delta_{k,1}\}}
\label{ZRCMans}
\ee
with
\be
\hat{\cal W}_1 = N_1N_2\,p_1+  (N_1+N_2)\sum_{a=1}^\infty ap_{a+1}\frac{\p}{\p p_a} +
 \sum_{a,b=1}^{\infty}\Big( (a+b-1)p_ap_b\frac{\p}{\p p_{a+b-1}}
+ab\,p_{a+b+1}\frac{\p^2}{\p p_a\p p_b}\Big)
\ee
These formulas for partition functions
were discovered and discussed in \cite{IMMrainbow,HMAMOsol},
but, in this section, we want to derive them from the $W$-representations.

The key is the generalization of (\ref{evW}), which is equivalent to
\be
\hat W_{\Delta} = \sum_R \varphi_{R,\Delta}\chi_R\{p\}\hat\chi_R
\ee
with the differential operator $\hat\chi_R = \chi_R\left\{k\frac{\p}{\p p_k}\right\}$,
which is a dual character, \cite{Mac}
\be\label{charort}
\left.\hat\chi_R \ \chi_{R'}\{p\}\right|_{p=0} = \delta_{R,R'}
\ee
In the particular case of (\ref{W2}) this means that
\be
\hat{  W}_{[2]} =
\frac{1}{2}\sum_{a,b=1}^\infty \Big((a+b)p_ap_b\p_{a+b} + ab\,p_{a+b}\p_a\p_b\Big)
=\sum_{|R|=|R'|} \alpha_{R,R'} \chi_R\{p\} \hat \chi_{R'}
= \chi_{[2]}\hat\chi_{[2]} - \chi_{[1,1]}\hat\chi_{[1,1]} + \nn\\
+(2\chi_{[3]}-\chi_{[2,1]})\cdot\hat\chi_{[3]} +(-\chi_{[3]}+ \chi_{[1,1,1]})\cdot\hat\chi_{[2,1]}
+ (\chi_{[2,1]}\hat\chi_{[1,1,1]} - 2\chi_{[1,1,1]})\cdot \hat\chi_{[1,1,1]}
+ \nn \\
+\ldots
\label{W2char}
\ee
As follows from (\ref{evW}), the operator $W_{[2]}$ has an eigenvalue $\varphi_{R,[2]}$, which is equal to
\be\label{varkappa}
\varphi_{R,[2]} = 2\sum_{(i,j)\in R}(j-i):=2\varkappa_R
\ee
Note that the operator at the r.h.s. of (\ref{W2char}) is {\it not} equal just to a diagonal sum $\sum_R \varkappa_R\chi_R\{p\}\hat \chi_R$ with the eigenvalue $\varkappa_R$.
This is because, before putting $p=0$, the orthogonality condition  (\ref{charort})
is not true, e.g. $\hat\chi_{[2]}\chi_{[3]}\{p\} = p_1$,
and
\be
\hat W_{[2]}\chi_{[3]}\{p\}  \ \stackrel{(\ref{W2char})}{=}\  \Big(\chi_{[2]}\hat\chi_{[2]}
+ (2\chi_{[3]}-\chi_{[21]})\chi_3 \Big)\chi_3 =
p_1\chi_{[2]} + 2\chi_{[3]}-\chi_{[2,1]} = 3\chi_{[3]} =\varkappa_{[3]}\chi_{[3]}
\ee

In general, lifting of (\ref{charort}) to the operator level
involves decomposition of the commutator into a sum of skew characters,
\be
 \hat\chi_R \cdot \chi_{R'}\{p\}    =
\sum_{Q}  \chi_{_{R'/Q}}\{p\}\cdot \hat\chi_{_{R/Q}}
\label{charortskew}
\ee
in particular,   picking up just the $c$-number piece at the r.h.s.,
we get the contribution from $Q=R$ only, i.e. $\hat\chi_{R} \cdot \chi_{R'}\{p\}=\chi_{R'/R}\{p\}$,
what is non-vanishing only for $|R'|\geq |R|$.

We remind that the skew characters are defined by the property
\be
\chi_R\{p+p'\} = \sum_{Q\subset R} \chi_{Q}\{p\} \chi_{R/Q}\{p'\}
\ee
and can be expressed via the usual Schur functions,
\be
\chi_{R/Q}\{p\} = \sum_{R'} C^{R}_{QR'}\chi_{R'}\{p\}
\ee
through the Littlewood-Richardson coefficients $C^{R}_{QR'}$, the structure constants
of the character multiplication
\be
\chi_R\{p\}\chi_{R'}\{p\} = \sum_{R''\in R\otimes R'} C^{R''}_{RR'} \chi_{R''}\{p\}
\ee
Coming back to (\ref{W2char}), the coefficient in front of $\hat\chi_Y$ is given
by a recursion formula
\be
\sum_{Y':\ |Y'|=|Y|}\alpha_{_{Y'Y}}\chi_{_{Y'}} \ = \
\varkappa_{_Y}\chi_{_Y} - \sum_{|R|=|R'|<|Y|}\alpha_{_{R,R'}}\chi_{_R}\chi_{_{Y/R'}}
\ee
One can calculate the coefficients $\alpha_{_{R,R'}}$ from this formula. Remarkably, these coefficients  are non-vanishing
only for the single-hook diagrams $R=[r,1^{s-1}]$ and $R'=[r',1^{s'-1}]$, and the final answer is
\be\boxed{
\frac{1}{2}\hat{  W}_{[2]} =
\frac{1}{2}\sum_{a,b=1}^\infty \Big((a+b)p_ap_b\p_{a+b} + ab\,p_{a+b}\p_a\p_b\Big)
= \sum_{\stackrel{r,s,r',s'=1}{r+s=r'+s' }}^\infty
(-)^{s+s'} (r-s') \cdot \chi_{_{[r,1^{s-1}]}}\cdot\hat\chi_{_{[r',1^{s'-1}]}}
}
\label{W2sh}
\ee
Note that despite only the single hook $\hat\chi$ contribute,
all $\chi_R$, not only single hook are   eigenfunctions of this $\frac{1}{2}\hat W_{[2]}$
with  non-vanishing eigenvalues $\varkappa_R$.
Note also that the operator at the l.h.s. of (\ref{W2sh}) contains at most second derivatives,
while particular items at the r.h.s. contain derivatives up to order $|R|$, though
all these higher derivatives cancel in the sum.

Now we want to do the same for ${\cal W}_2$ and ${\cal W}_1$, which
have a non-vanishing grading, thus the sums will not be diagonal even in the size
of the Young diagrams.
Instead,
\be
{\cal W}_2 =
N^2p_2+Np_1^2+2N\sum_{a=1}^\infty ap_{a+2}\p_a
+\sum_{a,b=1}^\infty \Big((a+b-2)p_ap_b\p_{a+b-2} + ab\,p_{a+b+2}\p_a\p_b\Big)
=\sum_{R\,\vdash |R'|+2} A_{R,R'} \chi_R\{p\} \hat \chi_{R'}
\nn
\ee
where the sum goes over the Young diagrams  $R$ and $R'$ which differ by $2$ in size.
Note that the operator at the l.h.s. contains at most second derivatives,
while particular items at the r.h.s. contain derivatives up to order $|R'|$.
Still all the higher derivatives cancel in the sum.
The coefficients $A_{R,R'}$ are linear functions of $N$,
and contributing are only the single-hook diagrams $R=[r,1^{s-1}]$ and $R'=[r',1^{s'-1}]$ so that the answer is
\be
\boxed{
\frac{1}{2}\hat{\cal W}_2
 =  \frac{1}{2}\Big((N+1)\chi_{[2]}-(N-1)\chi_{[1,1]}\Big)\cdot N+
\sum_{\stackrel{r,s,r',s'=1}{r+s=r'+s'+2}}^\infty
(-)^{s+s'} (N +r-s'-1)\cdot \chi_{_{[r,1^{s-1}]}}\cdot\hat\chi_{_{[r',1^{s'-1}]}}
}
\label{calW2}
\ee

Similarly, for $\hat {\cal W}_1 $,
contributing are only the single-hook diagrams $R=[r,1^{s-1}]$ and $R'=[r',1^{s'-1}]$,
and the answer is
\be
\boxed{
\hat {\cal W}_1 = \chi_{_{[1]}}\cdot N_1N_2 +
\sum_{\stackrel{r,s,r',s'=1}{r+s=r'+s'+1}}^\infty
(-)^{s+s'} (N_1+N_2+2r-2s'-1)\cdot \chi_{_{[r,1^{s-1}]}}\cdot\hat\chi_{_{[r',1^{s'-1}]}}
}
\label{W1expan}
\ee
This formula is invariant under simultaneous transposition of $R$ and $R'$,
accompanied by a sign inversion of $N_1$ and $N_2$.
Indeed, such transformation changes $(r,s,r',s')\longrightarrow(s,r,s',r')$,
thus $(-)^{s+s'}=(-)^{s-s'}\longrightarrow (-)^{r-r'} = (-)^{s'-s+1} = -(-)^{s-s'}$
and $2r-2s'-1\longrightarrow 2s-2r'-1 = -(2r-2s'-1)$, where in both cases we used
the constraint $r+s=r'+s'+1$.

Exponentiation of (\ref{calW2}) and (\ref{W1expan})
provides an unrestricted sum over $R$ and $R'$:
\be
e^{{\cal W}_2/2} = \sum_{R,R'} {\cal A}_{R,R'} \chi_R\{p\} \hat \chi_{R'}
\ee
with the coefficients ${\cal A}_{R,R'}$ non-zero only for $|R|-|R'|$ being even and non-negative.
Eq.(\ref{Zhermans}) is the piece of this sum with $R'=\emptyset$.
Note that items with $R$ and $R'$ of odd sizes in the expansion of ${\cal W}_2$
contribute to the exponential.
Likewise,  (\ref{ZRCMans}) is the piece of the sum
\be
e^{{\cal W}_1} = \sum_{R,R'} {\cal B}_{R,R'} \chi_R\{p\} \hat \chi_{R'}
\ee
with $R'=\emptyset$.

Exponentiation can be performed with the help of (\ref{charortskew}).
From (\ref{charortskew}) and (\ref{W1expan}), we obtain
\be
{\cal W}_1^2 = \!\!\!\!\! \sum_{R,R',R'',R'''}\!\!\! B_{R,R'}B_{R'',R'''} \chi_R\cdot\hat \chi_{R'}
\cdot \chi_{R'}\cdot\hat\chi_{R'''} = \!\!\!\!\!
\sum_{R,R',R'',R''',Q} \!\!\!\! B_{R,R'}B_{R'',R'''}
(\chi_R\cdot\chi_{R''/Q}) \cdot (\hat\chi_{R'/Q}\cdot\hat\chi_{R'''})
= \nn \\
= \!\!\!\!\!\!\!\!\!\!\!\!\!\!
\sum_{R,R',R'',R''',Q,Q',Q''} \!\!\!\!\!\!\!\!\!\!\!\!\!\!
B_{R,R'}B_{R'',R'''} C^{R''}_{QQ''} C^{R'}_{QQ'}
(\chi_R\cdot \chi_{Q''})\cdot(\hat \chi_{Q'}\cdot \hat\chi_{R''})
= \!\!\!\!\!\!\!\!\!\!\!\!\!\!
\sum_{R,R',R'',R''',Q,Q',Q''} \!\!\!\!\!\!\!\!\!\!\!\!\!\!
B_{R,R'}B_{R'',R'''} C^{R''}_{QQ''} C^{R'}_{QQ'}
C^{Y}_{RQ''}C^Z_{R''Q'}\cdot \chi_Y\cdot\hat \chi_Z
\nn
\ee
i.e.
\be
{\cal B}_{YZ}^{(2)} = \sum_{R,R',R'',R''',Q,Q',Q''} \!\!\!\!\!\!\!\!\!\!
B_{R,R'}B_{R'',R'''} C^{R''}_{QQ''} C^{R'}_{QQ'}
C^{Y}_{RQ''}C^Z_{R''Q'}
\ee
where ${\cal B}^{(2)}_{YZ}$ is the contribution from  ${\cal W}_1^2$
to the full matrix ${\cal B}_{YZ}$ for the exponential $e^{{\cal W}_1}$.
Diagrams of the type $R$ are all single-hook, thus the same is true for diagrams $Q$
appearing in the skew characters.
However, at the last stage the single-hook characters are multiplied,
and $Y$ and $Z$ are already 2-hook diagrams.
Likewise, the higher powers ${\cal W}_1^m$ involve transition matrices
${\cal B}_{YZ}^{(m)}$ between characters of $m$-hook diagrams.

\section{Unitary-type  (trigonometric)  models
\label{unita}}

When reduced to eigenvalues, the measure of Hermitian matrix model contains
a square of the Vandermonde determinant $\prod_{i<j} (m_i-m_j)^2$,
which has natural deformations and generalizations like
\be
\prod_{i<j} (m_i-m_j)^2 \longrightarrow
\prod_{i,j} \prod_{k=1}^{\beta} (m_i - q^{k-1}m_j)
\ee
and leads to substitution of the Schur functions by the Macdonald functions and their
various limits (like the Jack and Hall-Littlewood polynomials) \cite{Mac,MPS}.

There is, however, another important direction to generalize: to
a unitary\footnote{The trigonometric Vandermonde determinant has first emerged within the context of the unitary matrix models \cite{Umm}. However, the ``fair'' unitary model requires the choice of integration contour along the imaginary axis (in fact, on an imaginary segment) and averaged are bilinear combinations of characters. Hence, we refer to this type of models just as to trigonometric (though maybe more exact is the name ``hyperbolic''). Note that these models emerge most naturally as Chern-Simons type matrix models \cite{Mar}.}
or trigonometric model
\be
\prod_{i<j} (m_i-m_j)^2 \longrightarrow
\prod_{i<j} \sinh^2 \left(\frac{m_i-m_j}{2}\right)  \sim
\prod_{i<j} (m_i-m_j)^2\prod_n \left(1+ \frac{(m_i-m_j)^2}{4\pi^2 n^2}\right)^{\!2}
\ee
and, further, to the MacMahon  model:
\be
\longrightarrow
\prod_{i<j} (m_i-m_j)^2\prod_n \left(1+ \frac{(m_i-m_j)^2}{4\pi^2 n^2}\right)^{\!2n }
\label{McMahMAMO}
\ee
The former model was a testing area for initial studies of character expansions in matrix models
\cite{UniMAMO,AMMU,M2009}. It also emerged in the studies of localization of the ABJM theory \cite{KWY}.
The MacMahon model arises in the studies of localization of superconformal gauge theories \cite{Pest}
(there are also instanton correction, which vanish in the large $N$ \cite{Zarembo}
and other interesting limits).
They can be considered as ``perturbations'' of  Hermitian model by
bi-trace addition to the action
\be
2\cdot\sum_{i<j} \sum_n n^\nu \log \left(1+ \frac{(m_i-m_j)^2}{4\pi^2 n^2}\right)
=\sum_k \frac{(-)^{k+1} \zeta(2k-\nu)}{4^k\pi^{2k}k} \sum_{i,j} (m_i-m_j)^{2k}
= \nn \\
= \sum_{k=1}^\infty\sum_{i=0}^{2k} \frac{(-)^{k+i+1} \zeta(2k-\nu)}{4^k\pi^{2k}k} \cdot
\frac{(2k)!}{i!(2k-i)!}\cdot \Tr M^i\,\Tr M^{2k-i}
\label{bitrace}
\ee
with $\nu=0,1$ for trigonometric and MacMahon models respectively, and $\zeta(s)$ is the Riemann zeta-function.
In the former case, values of the $\zeta$-functions are elementary numbers (modulo powers of $\pi$),
while, in the latter case, they are transcendental,
but in existing applications this difference does not manifest itself.
This is because these theories are often studied perturbatively,
by expanding the exponentiated bi-trace into a series and taking averages
within the Hermitian matrix model.

However, it is much more interesting to look for {\it exact} formulas like (\ref{chardeco})
in the trigonometric model itself, without a reference to the Hermitian one.
For the trigonometric model, the statement is
\be
\boxed{
\left<  \chi_R[e^M] \right>^{\rm trig} =   A^{|R|}\cdot q^{2\varkappa_R } \cdot \chi_R\{p^*\}
}
\label{chardecouni}
\ee
where the average is taken with the weight
\be
\frac{1}{Z}
\int \prod_{i<j}^N \sinh^2\! \left(\frac{m_i-m_j}{2}\right)
\prod_{i=1}^N \exp\left({-\frac{m_i^2}{2g^2}}\right) dm_i
\label{chardecounint}
\ee
When averaging, the argument of character in the integrand is the diagonal matrix
with the entries $e^{m_i}$,
and, at the r.h.s., the parameters are
$q = e^{g^2/2}$, $A=q^N=e^{Ng^2/2}$, the exponent $\varkappa_R = \sum_{(x,y)\in R}(y-x)$ is the same as in (\ref{varkappa}). The time variables $p^*_k = \frac{A^{k}-A^{-k}}{q^k-q^{-k}}$ in the argument
of the character at the r.h.s. lie in the
``topological locus'' obtained by the $q$-deformation of $p_k = N$.
Thus, the difference from (\ref{chardeco}) are a quantum deformation, and
the drastic change of the combinatorial factor from the ratio of characters
to the exponential of the eigenvalue of the second Casimir operator in the representation $R$, $\varkappa_R$.
The limit $q\longrightarrow 1$ is trivial: it corresponds to $g^2\longrightarrow 0$,
when the Gaussian exponential $e^{\Tr M^2/2g^2}$ turns into the $\delta$-function
so that the integral in (\ref{chardecounint}) gets localised at $M=0$, i.e. $U=I_N$,
and this relations just gives rise to the identity $\chi_R[I_N] = \chi_R\{N\}$.
In the opposite limit of $g^2\longrightarrow \infty$, the Gaussian exponential disappears,
so the measure reduces to the Haar measure, but the integral diverges, and so does the r.h.s.
of  (\ref{chardecouni}), where $q\longrightarrow \infty$.
One could instead consider true unitary integrals with pure imaginary $m_i$ and
unimodular $q$, but, in this case, non-vanishing are only balanced averages,
with the same number of $U$ and $U^\dagger$, which now differs from $U$
and rather equal to $U^{-1}$. This is a more interesting and complicated case,
with the 't Hooft - de Wit anomalies and other peculiarities, see \cite{UniMAMO,AMMU}
and references therein.

\bigskip

We can now compare the implications of (\ref{bitrace}) at $\nu=0$ with those of
(\ref{chardecouni}).
As the simplest example consider
\be
\left< \chi_1[e^M]\right>^{\rm trig} \stackrel{(\ref{chardecouni})}{=}
 \frac{e^{g^2N}-1}{e^{g^2/2}-e^{-g^2/2}}
= N + \frac{g^2N^2}{2} + \frac{g^4N(4N^2-1)}{24}+\frac{g^6N^2(2N^2-1)}{48} + \ldots
\label{avunitcgi1}
\ee
On the other hand, the same quantity is just the ratio of the Hermitian model averages
\be
\left< \chi_1[e^M]\right>^{\rm trig} \stackrel{(\ref{bitrace})}{=}
\frac{\left<\exp\left(\sum_{k,i} \frac{(-)^{k+i+1} \zeta(2k)}{(2\pi)^{2k}k} \cdot
\frac{(2k)!}{i!(2k-i)!}\cdot \Tr M^i\,\Tr M^{2k-i} \right)\cdot \Tr e^M     \right>}
{\left< \exp\left(\sum_{k,i} \frac{(-)^{k+i+1} \zeta(2k)}{(2\pi)^{2k}k} \cdot
\frac{(2k)!}{i!(2k-i)!}\cdot \Tr M^i\,\Tr M^{2k-i} \right)
\right> }
= \nn \\
= \frac{ \left<
\left(1+ \sum_{k,i} \frac{(-)^{k+i+1} \zeta(2k)}{(2\pi)^{2k}k} \cdot
\frac{(2k)!}{i!(2k-i)!}\cdot \Tr M^i\,\Tr M^{2k-i}
+ \ldots\right)
\left(N + \frac{1}{2}\Tr M^2 + \frac{1}{24}\Tr M^4 + \ldots\right)
\right>}{\left<
1+ \sum_{k,i} \frac{(-)^{k+i+1} \zeta(2k)}{(2\pi)^{2k}k} \cdot
\frac{(2k)!}{i!(2k-i)!}\cdot \Tr M^i\,\Tr M^{2k-i} +
\frac{1}{2}\left(\sum_{k,i} \frac{(-)^{k+i} \zeta(2k)}{(2\pi)^{2k}k} \cdot
\frac{(2k)!}{i!(2k-i)!}\cdot \Tr M^i\,\Tr M^{2k-i}\right)^2 + \ldots
\right>}
= \nn
\ee
{\footnotesize
\be
= N + \frac{1}{2}<\Tr M^2> + \frac{1}{24}<\Tr M^4> + \left<\frac{1}{2}\Tr M^2
\cdot \frac{2\zeta(2)}{\pi^2}\Big(N\Tr M^2 - (\Tr M)^2\Big)\right>
- \left<\frac{1}{2}\Tr M^2\right>
\cdot \left<\frac{2\zeta(2)}{\pi^2}\Big(N\Tr M^2 - (\Tr M)^2\Big)\right> + O(g^6)
= \nn \\
= N + \frac{g^2N^2}{2}+ g^4\left\{\frac{N(2N^2+1)}{24}
+ \frac{\zeta(2)}{4\pi^2}\Big(N^3(N^2+2)-N(N^2+2) - N^2(N^3 - N)\Big)\right\}
+ \ldots
= N + \frac{g^2N^2}{2}+ \frac{g^4N(4N^2-1)}{24} + O(g^6)
\nn
\ee
}

\noindent
and, substituting $\zeta(2)=\frac{\pi^2}{6}$, we reproduce (\ref{avunitcgi1}).
Already this simple example clearly demonstrates the advantage,
even technical of exact formulas
like (\ref{chardecouni}) over the perturbation expansions like (\ref{bitrace}).

\section{Knot matrix models}

There is another interesting way to deform the trigonometric model \cite{BEMT}:
\be
\prod_{i<j} \sinh^2 \left(\frac{m_i-m_j}{2}\right) \longrightarrow
\prod_{i<j} \sinh \frac{m_i-m_j}{2a} \prod_{i<j} \sinh\frac{m_i-m_j}{2b}
\ee
Like the trigonometric model (\ref{chardecounint}), it also preserves characters,
moreover, (\ref{chardecouni}) remains true,
\be
\boxed{
\left< \chi_R[e^{M/a}] \right>^{[a,b]}
=   \left(A^{|R|}\cdot q^{2\varkappa_R }\right)^{b/a} \cdot \chi_R\{p^*\}
}
\label{abchardeco}
\ee
what changes is only the value of $q=\exp{\frac{g^2}{2ab}}$.
This formula "spontaneously breaks" the $a\leftrightarrow b$ symmetry,
and has the corresponding counterpart
\be
\left< \chi_R[e^{M/b}] \right>^{[a,b]}
=   \left(A^{|R|}\cdot q^{2\varkappa_R }\right)^{a/b} \cdot \chi_R\{p^*\}
\label{abchardeco1}
\ee
with the same $q$.

This measure appears in description of the HOMFLY polynomials
for torus knots with $a$ and $b$ coprime:
\be\label{52}
\left< \chi_R[e^{M}] \right>^{[a,b]}
 = {\cal H}_R^{{\rm Torus}_{a,b}}(A,q)
 = D_R(N) \cdot H_R^{{\rm Torus}_{a,b}}(A,q)
\ee
The character at the l.h.s.  depends on $e^M$ and, before (\ref{abchardeco}) can be applied,
one needs to express it through characters of $e^{M/a}$.
This decomposition involves a combination of characters for Young diagrams
of the size $a|R|$ with peculiar Adams coefficients $c$.
After this, substitution (\ref{abchardeco}) converts (52) into the Rosso-Jones formula \cite{RJ}:
\be
{\cal H}_R^{{\rm Torus}_{a,b}}(A,q) =
\left< \chi_R[e^{M}] \right>^{[a,b]} =
\left< \sum_{Q\vdash a|R|} c_{R,Q} \chi_Q[e^{M/a}] \right>^{[a,b]}
= A^{\frac{b|R|}{a}}\sum_{Q\vdash a|R|} c_{R,Q}
\cdot q^{\frac{2b \varkappa_Q}{a} } \cdot \chi_Q\{p^*\}
\ee
In result, $H_R^{{\rm Torus}_{a,b}}$  is a Laurent polynomial of $q$ and $A$
(it remains such for arbitrary knots, not only torus).
Within this framework, the equivalence of (\ref{abchardeco}) and (\ref{abchardeco1})  reflects the
Reidemeister equivalence of $a$-strand and $b$-strand realizations of the
same torus knot ${\rm Torus}_{[a,b]}$.
For  attempts to extend matrix model description beyond the torus knots
see \cite{AMMMknotMAMO}.

When $a$ and $b$ have a non-trivial common divisor,
we get a torus {\it link} instead of a knot,
and its HOMFLY invariant is an average of a product of characters,
as many as there are components in the link (the number equal to the common divisor of $a$ and $b$).
When $a=b=2$, this is actually the Hopf link, and its HOMFLY invariant is again a character,
see \cite{MMHopf} for a review and references.
At the same time, the measure in this case is exactly that of the trigonometric model.
In other words, we conclude that
\be\label{dH}
q^{-2\varkappa_R-2\varkappa_S}A^{-|R|-|S|}\cdot\Big< \chi_R[e^{M/2}]\,\chi_S [e^{M/2}]\Big>^{{\rm trig}} =
{\cal H}_{R\times S}^{\rm Hopf} = D_R(N) \chi_S\{p^{*R}\} = D_S(N)\chi_R\{p^{*S}\}
\ee
for
\be\label{t*}
p_k^{*R} = \frac{q^{Nk}-q^{-Nk}}{q^k-q^{-k}}
+ \sum_{i=1}^{l_R}q^{k(N-2i+1)}\Big(q^{2kr_i}-1\Big)
\ee
The framing factor at the l.h.s. of (\ref{dH}) has to be taken in degree ${1\over 2}\Big({a\over b}+{b\over a}\Big)$ in the generic torus knot/link case.

This strangely-looking shift of time variables (\ref{t*}) is in fact induced by action
of the cut-and-join operator on the "topological locus"
$p_k^* = \frac{q^{Nk}-q^{-Nk}}{q^k-q^{-k}}$,
which is the trigonometric-model substitute of the locus $p_k=N$ in the Hermitian models:
\be
 \left.e^{2\hat W_{[2]}} \chi_R\{\bar p\}\chi_S\{\bar p\}\right|_{
\bar p_k = p_k^*}=q^{2\varkappa_R+2\varkappa_S}D_R(N)\chi_S\{p_k^{*R}\}
\ee

In the case of torus measure (\ref{abchardeco}) with arbitrary $a$ and $b$, the
bi-trace correction to the action (\ref{bitrace}) is substituted by
\be
\frac{1}{2}\sum_{i,j} \sum_n  \left\{ \log \left(1
+  \frac{1}{a^2}\frac{(m_i-m_j)^2}{4\pi^2 n^2}\right)
+ \log \left(1
+  \frac{1}{b^2}\frac{(m_i-m_j)^2}{4\pi^2 n^2}\right)\right\}
=\sum_k \frac{(-)^{k+1} \eta_k\zeta(2k)}{4^k\pi^{2k}k} \sum_{i,j} (m_i-m_j)^{2k}
= \nn \\
= \sum_{k=1}^\infty\sum_{i=0}^{2k} \frac{(-)^{k+i+1} \eta_k\zeta(2k)}{4^k\pi^{2k}k} \cdot
\frac{(2k)!}{i!(2k-i)!}\cdot \Tr M^i\,\Tr M^{2k-i}
\label{bitraceab}
\ee
with $\nu=0$ and $\eta_k=\frac{1}{2}\left(\frac{1}{a^{2k}}+\frac{1}{b^{2k}}\right)$. Then
\be
\left< \chi_1[e^{M/a}]\right>^{[a,b]} \stackrel{(\ref{bitraceab})}{=}
\frac{\left<\exp\left(\sum_{k,i} \frac{(-)^{k+i+1}\eta_k \zeta(2k)}{(2\pi)^{2k}k} \cdot
\frac{(2k)!}{i!(2k-i)!}\cdot \Tr M^i\,\Tr M^{2k-i} \right)\cdot \Tr e^{M/a}     \right>}
{\left< \exp\left(\sum_{k,i} \frac{(-)^{k+i+1} \eta_k\zeta(2k)}{(2\pi)^{2k}k} \cdot
\frac{(2k)!}{i!(2k-i)!}\cdot \Tr M^i\,\Tr M^{2k-i} \right)
\right> }
= \nn \\
= \frac{ \left<
\left(1+ \sum_{k,i} \frac{(-)^{k+i+1} \eta_k\zeta(2k)}{(2\pi)^{2k}k} \cdot
\frac{(2k)!}{i!(2k-i)!}\cdot \Tr M^i\,\Tr M^{2k-i}
+ \ldots\right)
\left(N + \frac{1}{2a^2}\Tr M^2 + \frac{1}{24a^4}\Tr M^4 + \ldots\right)
\right>}{\left<
1+ \sum_{k,i} \frac{(-)^{k+i+1} \eta_k\zeta(2k)}{(2\pi)^{2k}k} \cdot
\frac{(2k)!}{i!(2k-i)!}\cdot \Tr M^i\,\Tr M^{2k-i} +
\frac{1}{2}\left(\sum_{k,i} \frac{(-)^{k+i} \eta_k\zeta(2k)}{(2\pi)^{2k}k} \cdot
\frac{(2k)!}{i!(2k-i)!}\cdot \Tr M^i\,\Tr M^{2k-i}\right)^2 + \ldots
\right>}
= \nn
\ee
{\footnotesize
\be
\!\!\!\!\!\!\!\!\!\!\!\!\!\!\!= N + \frac{1}{2a^2}<\Tr M^2> + \frac{1}{24a^4}<\Tr M^4> + \left<\frac{1}{2a^2}\Tr M^2
\cdot \frac{2\eta\zeta(2)}{\pi^2}\Big(N\Tr M^2 - (\Tr M)^2\Big)\right>
- \left<\frac{1}{2a^2}\Tr M^2\right>
\cdot \left<\frac{2\eta_1\zeta(2)}{\pi^2}\Big(N\Tr M^2 - (\Tr M)^2\Big)\right> + O(g^6)
=\nn
\ee
}
\be
= N + \frac{g^2N^2}{2a^2}+ g^4\left\{\frac{N(2N^2+1)}{24a^4}
+ \frac{\eta_1\zeta(2)}{4\pi^2a^2}\Big(N^3(N^2+2)-N(N^2+2) - N^2(N^3 - N)\Big)\right\}
+ \ldots
= \nn \\
= N + \frac{g^2N^2}{2a^2}
+ \frac{g^4N }{24a^4}\left(2N^2+1+\frac{12a^2\eta_1\zeta(2)}{\pi^2}(N^2-1)\right) + O(g^6)
= \nn \\
= e^{\frac{g^2N}{2a^2}}\cdot N  \left\{1 +
\frac{g^4 }{24a^4}\underbrace{\left(\frac{12a^2\eta_1\zeta(2)}{\pi^2}-1\right)}_{a^2/b^2}(N^2-1)
+ O(g^6)\right\}
= e^{\frac{g^2N}{2a^2}}\cdot \frac{\sinh\frac{g^2N}{2ab}}{\sinh\frac{g^2}{2ab}}
\ee
A generalization of the torus matrix model to non-torus knots
is an open problem: for $N=2$ it is nicely solved in \cite{AMMknotMAMO},
but the matrix model lifting to arbitrary $N$ remains a challenge.

\section{Restriction to traceless matrices}

Let us start with the Hermitian matrix model (\ref{chardeco}). The restriction can be imposed in different ways.
The simplest is just to insert a $\delta$-function in the form
$\delta(\Tr M) = \frac{1}{2\pi}\int e^{i\alpha\Tr M} d\alpha$.
This is equivalent to shifting the integration variable $M\longrightarrow M+i\alpha$
and integrating the answer over $\alpha$ with the Gaussian measure
$\exp\Big(-\frac{N\alpha^2}{2}\Big) d\alpha$:
\be
\Big<F(M)\Big>^{\rm traceless} =
\sqrt{N\over 2\pi}
\int \Big<F(M+i\alpha)\Big>\cdot e^{-\frac{N\alpha^2}{2}}d\alpha
\ee
For example,
\be
\Big< (\Tr M)^2 \Big>^{\rm traceless} =
\sqrt{N\over 2\pi}
\int \Big< (\Tr M)^2 -N^2\alpha^2 \Big> e^{-\frac{N\alpha^2}{2g^2}}d\alpha
= N-N=0
\ee
while
\be
\Big< \Tr M^2 \Big>^{\rm traceless} =
\sqrt{N\over 2\pi}
\int \Big< \Tr M^2 -N\alpha^2 \Big> e^{-\frac{N\alpha^2}{2g^2}}d\alpha
= N^2-1
\ee
so that $\Big< \chi_{[2]}[M] \Big>^{\rm traceless} = \frac{(N+1)(N-1)}{2}$
and   $\Big< \chi_{[1,1]}[M] \Big>^{\rm traceless} = -\frac{(N+1)(N-1)}{2}$.

In the generic case,
\be
\Big<\chi_R[M]\Big>^{\rm traceless} =
\left(\frac{ N}{2\pi}\right)^{1/2}
\int \Big<\chi_R[M+i\alpha I]\Big> \cdot e^{-\frac{N\alpha^2}{2}}d\alpha =
\nn \\
=\sum_{S\subset R}
\frac{\chi_{R/S}\{\delta_{k,1}\}\chi_S\{\delta_{k,1}\}}{\chi_R\{\delta_{k,1}\}}
\cdot \frac{D_R(N)}{D_S(N)}\cdot\Big<\chi_S[M ]\Big> \cdot
\left(\frac{N}{2\pi}\right)^{1/2}
\int(i\alpha)^{|R|-|S|}e^{-\frac{N\alpha^2}{2}}d\alpha
\ee
Only $R$ and $S$ of even sizes contribute to the sum. The integral of $\alpha$ is very immediate,
\be
<\alpha^{2k}>\ = \sqrt{2\pi\over N}\frac{(2k-1)!!}{N^{k}}
\ee
so that finally
\be\boxed{
\Big<\chi_R[M]\Big>^{\rm traceless} 
=\sum_{S\subset R}
i^{\frac{|R|-|S|}{2}}\cdot{(|R|-|S|-1)!!\over N^{\frac{|R|-|S|}{2}}} \cdot
\frac{\chi_{R/S}\{\delta_{k,1}\}\chi_S\{\delta_{k,1}\}}{\chi_R\{\delta_{k,1}\}}
\cdot \frac{D_R(N)}{D_S(N)}\cdot\Big<\chi_S[M ]\Big>}
\ee
The sum $\sum_{S\vdash s}
\frac{\chi_{R/S}\{\delta_{k,1}\}\chi_S\{\delta_{k,1}\}}{\chi_R\{\delta_{k,1}\}}= \frac{|R|!}{s!(|R|-s)!}$,
but the individual coefficients contain skew characters and thus are a little more involved.
For example, 
\be
\Big<\chi_{[2,2]}\Big>^{\rm traceless} \longrightarrow
\Big<\chi_{[2,2]}\Big> - {3\over N}\frac{D_{[3,1]}}{D_{[2]}}\Big<\chi_{[2]}\Big>
- {3\over N}\frac{D_{[2,2]}}{D_{[1,1]}}\Big<\chi_{[1,1]}\Big>
+{3\over N^2} \alpha^4 D_{[2,2]} 
\ee

For the trigonometric models including the MacMahon one, the restriction to the
traceless matrices works much simpler:
since $\chi_R[e^{M+i\alpha}] = e^{i\alpha|R|}\chi_R[e^M]$,
the $\alpha$-dependence factors out, and its only effect is
the additional factor  $\left(\frac{ N}{2\pi}\right)^{1/2}
\int e^{i\alpha |R|} \cdot e^{-\frac{N\alpha^2}{2}}d\alpha\ = e^{-\frac{-|R|^2}{2N}}$
in the average:
\be
\boxed{
\Big<\chi_R[e^{M/a}]\Big>^{\rm traceless}
= e^{-\frac{ |R|^2}{2a^2N}}\cdot \Big<\chi_R[e^{M/a}]\Big>
}
\label{unitraceless}
\ee
In this formula, the average can be taken at any model: Hermitian,
trigonometric, toric, since the generalized Vandermonde factors in the measure
do not depend on $\alpha$.

\section{Conclusion}

The central formula of this paper,
\be
\Big<\chi_R\{\Tr M^k\}\Big> \sim \chi_R\{p_k=N\}
\label{hermave}
\ee
{\it looks like} a statement that integration over $M$ is reduced
to the substitution of the "mean field" $M=Id$.
This would look mysterious, but in fact this is {\it not quite true}:
for an arbitrary function $F\{p_k\}$
\be
\Big<F\{\Tr M^k\}\Big>\ \ /\!\!\!\!\!\!\sim \ F\{p_k=N\}
\ee
the property is true only for the characters,
and it is another kind of a mystery,
more similar to a Duistermaat-Heckman (localization) trick
with group theory origins rather than to any kind of
an ordinary mean field calculation in quantum field theory.

In the trigonometric case, the situation is  different:
\be
\Big<\chi_R\{\Tr e^{kM}\}\Big>^{\rm trig}
\sim \chi_R\left\{p_k=\frac{\sinh(kNg^2/2)}{\sinh(kg^2/2)}\right\}
\ee
It {\it does not look like} a mean-field  formula:
the exponentials at the l.h.s. turn into
a somewhat different structure, the ratio of sinh's at the r.h.s.
Instead, it is nicely consistent with the quasiclassical approximation $\mu\to \infty$ in (\ref{chardeco}):
then dominating in the integral over $M$ is the vicinity of $M=0$
where $\Tr e^{kM} = N$.
Note that, in the Hermitian case, taking the limit $\mu\to\infty$
makes no sense: the $\mu$-dependence is fixed by dimensions of the operators:
there is no any weak coupling regime at all,
and formulas like (\ref{hermave}) are exact.

For straightforward $q,t$-deformation  of (\ref{hermave}) see \cite{MPS},
the clever thing to do in this case is just to take (\ref{hermave})
as a {\it definition} of the model, which is much simpler and more practical
than to proceed through multiple Jackson integrals and Pochhammer symbols.

Challenging are generalizations of (\ref{hermave}) in at least five directions:
\begin{itemize}
\item to non-Gaussian phases, where changing is only the coefficient in front of the
characters at the r..h.s., see \cite{PSh} for simplest examples,
\item to generic knots, not only torus ones, for a more detailed description of the problem
see \cite{AMMMknotMAMO},
\item to the MacMahon matrix model (\ref{McMahMAMO}),
where the Vandermonde determinant is substituted by a product of
the Barnes double $\Gamma$-functions,
\item to the network models \cite{MMZ,DIM} describing contractions of multiple
topological vertices, usual and refined,
\item to the Aristotelian tensor models of \cite{tenmod,Rang,IMMarist}.
\end{itemize}

The challenge is well illustrated already by the operator counting rules.
The ordinary characters (Schur functions) and their MacDonald deformations
are labeled by the ordinary Young diagrams, and their abundance
is described by the generating function
\be
\sum_n \#_{\rm Young}\cdot q^n = \prod_{n=1} \frac{1}{1-q^n}
\ee
For the plain partitions, which are labeling representations of the DIM algebra and the affine Yangian \cite{DIMaY}, it becomes
\be
\sum_n \#_{\rm plain}\cdot q^n =  \prod_{n=1} \frac{1}{(1-q^n)^{^n}}
\ee
while the number of gauge invariant operators in the Aristotelian
(rang 3 rainbow) tensor model grows even faster \cite{Rang,IMMarist}:
\be
\sum_n \#_{\rm Arist}\cdot q^n =
\prod_{n=1} \frac{1}{(1-q)(1-q^2)^{^3}(1-q^3)^{^7}(1-q^4)^{^{26}}(1-q^5)^{^{97}}
(1-q^6)^{^{624}}(1-q^7)^{^{4163}}\ldots}=\prod_{n=1}{1\over (1-q^n)^{\beta_n}}
\ee
where $\beta_n$ is the number of unlabeled dessins d'enfants with $n$ edges \cite{Gro}.

\bigskip

Our main purpose in this paper was to demonstrate a technical possibility
to attack all these problems in  a systematic way.
Our main emphasize was on the way the character-preservation property unifies
highly non-trivial and even previously unnoticed identities (sum rules) between
characters.
Usually such non-linear relations are described in terms of ``integrability",
but physically relevant quantities (non-perturbative partition functions)
are long known to be {\it more} than integrable,
the word {\it superintegrable}  seems most adequate to describe the situation.
In the standard language of matrix models, the story is that {\it matrix model
$\tau$-functions} are not just $tau$-functions, but satisfy an additional
{\it string equation} (and, in result, the whole set of Virasoro or $W$- constraints),
which altogether makes the model not just integrable, but {\it explicitly solvable}
like additional integrals of motion do for superintegrable mechanical systems.
However, the higher symmetry behind the superintegrable models, more complicated
than the Coulomb force with its hidden $O(d+1)$ symmetry is still under investigated
and is not straightforward to reveal, because it is non-linearly realized.
This paper can be considered as a substantial step in this direction,
which is based on the technique of character decompositions \cite{Char,MKR10,tenmod,HMAMOsol,IMMarist}, see earlier reviews in \cite{M2009}.
One should now study character decompositions of various harmonics of higher $W$-operators,
including the higher cut-and-join operators $W_R$ of \cite{MMN}, which are their zero-harmonics.
These will again involve new summation rules for characters, which can, however,
be more comprehensive than those implied by the genus expansions.
One of the issues is to study the emerging hook structure of the sum rules (hook formulas)
and its dependence on the shape of the diagram $R$.
Exponentiation of cut-and-join operators,
which is a kind of trivial since all the eigenfunctions and eigenvalues
are explicitly known from \cite{MMN}
is, however, also a source of highly non-trivial sum rules for characters.
To conclude, the character-preservation property of matrix models reveals
an entire new world of non-linear relations between the characters of linear and symmetric groups,
which requires an understanding from the point of view of the basic group theory.
This is especially important, because the combinatorial solution of matrix models
survives various deformations: from Young diagrams to plain partitions,  from matrices to tensors,
from the Gaussian to higher Airy measures,
from the Hermitian to trigonometric model and, probably, further, 
while the corresponding deformations of Lie algebra theory are yet unknown or, at best,
extremely complicated.
As usual, the matrix model approach provides a unifying view on the full set of problems
and provides an efficient method to solve them.

\section*{Acknowledgements}

Our work was supported by the Russian Science Foundation (Grant No.16-12-10344).

\end{document}